\documentclass[twocolumn,secnumarabic,amssymb, nobibnotes, aps, prl]{revtex4-2}
 
\usepackage[utf8]{inputenc}
\usepackage{graphicx}
\usepackage{bm}
\usepackage[dvipsnames]{xcolor}
\usepackage[normalem]{ulem}
\usepackage{textcomp}
\usepackage{hyperref}
\usepackage{hyperref}
\usepackage{xspace}
\usepackage{siunitx}
\usepackage{times}
\usepackage{amsmath}
\usepackage{amssymb}
\usepackage{svg}
\usepackage{upgreek}
\newcommand{\WTe}{WTe$_2$\xspace}

\newcommand{\NbSe}{NbSe$_2$\xspace}
\newcommand{\didv}{{$dI/dV$}\,}

\newcommand{\EF}{$E_{\rm F}$\xspace}

\begin{document}

\title{One-dimensional topological superconductivity in a van der Waals heterostructure}

	\author{
		Jose Martinez-Castro,$^{1,2,3,\ast,\dagger}$
		Tobias Wichmann,$^{1,2,4,\dagger}$
		Keda Jin,$^{1,2,3}$
		Tomas Samuely,$^{5}$
		Zhongkui Lyu,$^{1,2,4}$
		Jiaqiang Yan,$^{6}$
		Oleksander Onufriienko,$^{5}$
		Pavol Szab\'o,$^{5}$ 
		F. Stefan Tautz,$^{1,2,4}$
		Markus Ternes,$^{1,2,3}$
		and Felix L\"upke$^{1,2}$\\
		\vspace{0.25cm}
		\normalsize{$^{1}$\textit{Peter Gr\"unberg Institut (PGI-3), Forschungszentrum J\"ulich, 52425 J\"ulich, Germany}}\\
		\normalsize{$^{2}$\textit{J\"ulich Aachen Research Alliance, Fundamentals of Future Information Technology, 52425 J\"ulich, Germany}}\\
		\normalsize{$^{3}$\textit{Institut f\"ur Experimentalphysik II B, RWTH Aachen, 52074 Aachen, Germany}}\\
		\normalsize{$^{4}$\textit{Institut f\"ur Experimentalphysik IV A, RWTH Aachen, 52074 Aachen, Germany.}}\\
		\normalsize{$^{5}$\textit{Centre of Low Temperature Physics, Faculty of Science, Pavol Jozef Šafárik University \& Institute of Experimental Physics, Slovak Academy of Sciences, 04001 Košice, Slovakia}}\\
		\normalsize{$^{6}$\textit{Materials Science and Technology Division, Oak Ridge National Laboratory, Oak Ridge, TN 37831, USA}}\\
		\vspace{0.25cm}
		\normalsize{$^\dagger$These authors contributed equally}\\
		\normalsize{$^\ast$E-mail: j.martinez@fz-juelich.de}\\
	}

\maketitle

\textbf{
One-dimensional (1D) topological superconductivity is a state of matter that is not found in nature. However, it can be realised, for example, by inducing superconductivity into the quantum spin Hall edge state of a two-dimensional topological insulator \cite{Bocquillon2016, Jack2019}. Because topological superconductors are proposed to host Majorana zero modes \cite{Kitaev2001,Alicea2012,Flensberg2021}, they have been suggested as a platform for topological quantum computing \cite{Fu2008, Maeno2012}. Yet, conclusive proof of 1D topological superconductivity has remained elusive \cite{Bocquillon2016, Liu2017,Liu2018, Prada2020,Dartiailh2021,das2023}. Here, we employ low-temperature scanning tunnelling microscopy to show 1D topological superconductivity in a van der Waals heterostructure by directly probing its superconducting properties, instead of relying on the observation of Majorana zero modes at its boundary \cite{Nadj-Perge2014,Jeon2017,Kim2018, kezilebieke2020}. We realise this by placing the two-dimensional topological insulator monolayer \WTe on the superconductor \NbSe. We find that the superconducting topological edge state is robust against magnetic fields, a hallmark of its triplet pairing. Its topological protection is underpinned by a lateral self-proximity effect, which is resilient against disorder in the monolayer edge. By creating this exotic state in a van der Waals heterostructure, we provide an adaptable platform for the future realization of Majorana bound states. Finally, our results more generally demonstrate the power of Abrikosov vortices as effective experimental probes for superconductivity in nanostructures.\\
}

\begin{figure*}[!htbp]
  \centering
  \includegraphics[width=1\textwidth]{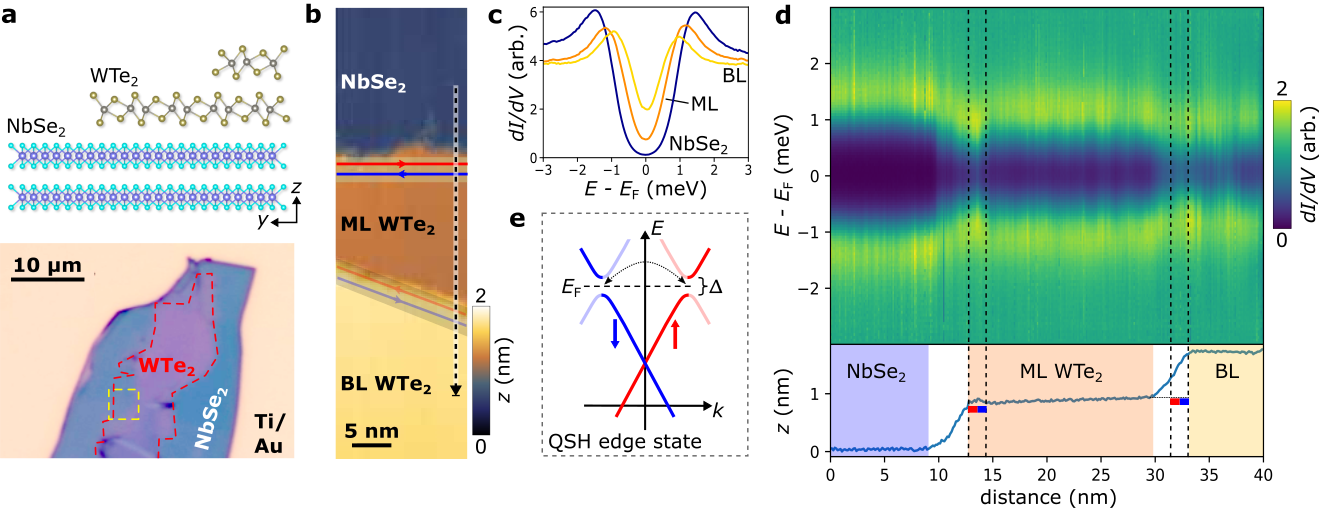}
  \caption{\label{Fig1} \textbf{Proximity-induced superconductivity in a \WTe/\NbSe heterostructure.}
  \textbf{(a)} Cross-sectional atomic structure model (top) and optical micrograph (bottom) of the heterostructure. The  outline of the \WTe flake is indicated by a dashed red line, the investigated area by a yellow dashed line. \textbf{(b)} STM topography of the \WTe flake edge.  The QSH edge state around the monolayer (ML) region is indicated by red and blue arrows. \textbf{(c)} \didv spectra of \NbSe, monolayer (ML) and bilayer (BL) \WTe, each showing the respective superconducting pairing gap ($T_{\rm eff}\approx2\rm\,K$). \textbf{(d)} Spectral map (top) taken along the dashed arrow in panel b and corresponding topography cross section (bottom) with QSH edge states indicated in red and blue. Edge state regions are marked by dashed black lines in the spectral map. \textbf{(e)} Schematic of the ML \WTe QSH edge state band structure. Inducing a supercondcuting pairing gap $\Delta$ at the Fermi energy \EF gives rise to an effective 1D $p$-wave superconductor. Faint lines indicate the particle-hole symmetric spectrum.}
\end{figure*}

Intrinsic one-dimensional (1D) superconductivity is unattainable because in one dimension quantum fluctuations suppress long-range order \cite{hohenberg1967}. This fundamental limitation can be overcome by inducing superconductivity via the proximity effect into quasi-1D systems such as nanowires \cite{doh2005}, magnetic atomic chains \cite{nadj2014}, topological boundary states of quantum spin Hall (QSH) insulators \cite{Hart2014,Bocquillon2016}, or higher-order topological insulators \cite{choi2020}. Topological edge states in van der Waals (vdW) heterostructures are a natural platform to engineer 1D topological superconductivity (TSC), because different materials can be stacked easily and with atomically clean interfaces. Moreover, the surfaces of vdW heterostructures are accessible to scanning tunnelling microscopy (STM), in contrast to buried quantum well structures. This facilitates local spectroscopy which also has been an effective tool in the search for Majorana states in 2D topological superconductors \cite{wang2018, kezilebieke2020,Nayak2021,Li2022}. A prototypical material to realize 1D TSC is  monolayer (ML) WTe$_2$, an intrinsic QSH insulator \cite{Qian2014, Crommie2017,Jia2017,Shi2019}. The spin-polarized edge states in ML \WTe show quantized transport \cite{Wu2018, Fatemi2018, zhao2021} and behave like a Tomonaga-Luttinger liquid \cite{jia2022}, which establishes their one-dimensional metallicity. Although the spectroscopic features of the QSH edge state coexist with an induced superconducting gap in ML \WTe/\NbSe heterostructures \cite{lupke2020, Weber2022}, this does not yet prove 1D TSC. 

\begin{figure*}[!htbp]
  \centering
  \includegraphics[width=1\textwidth]{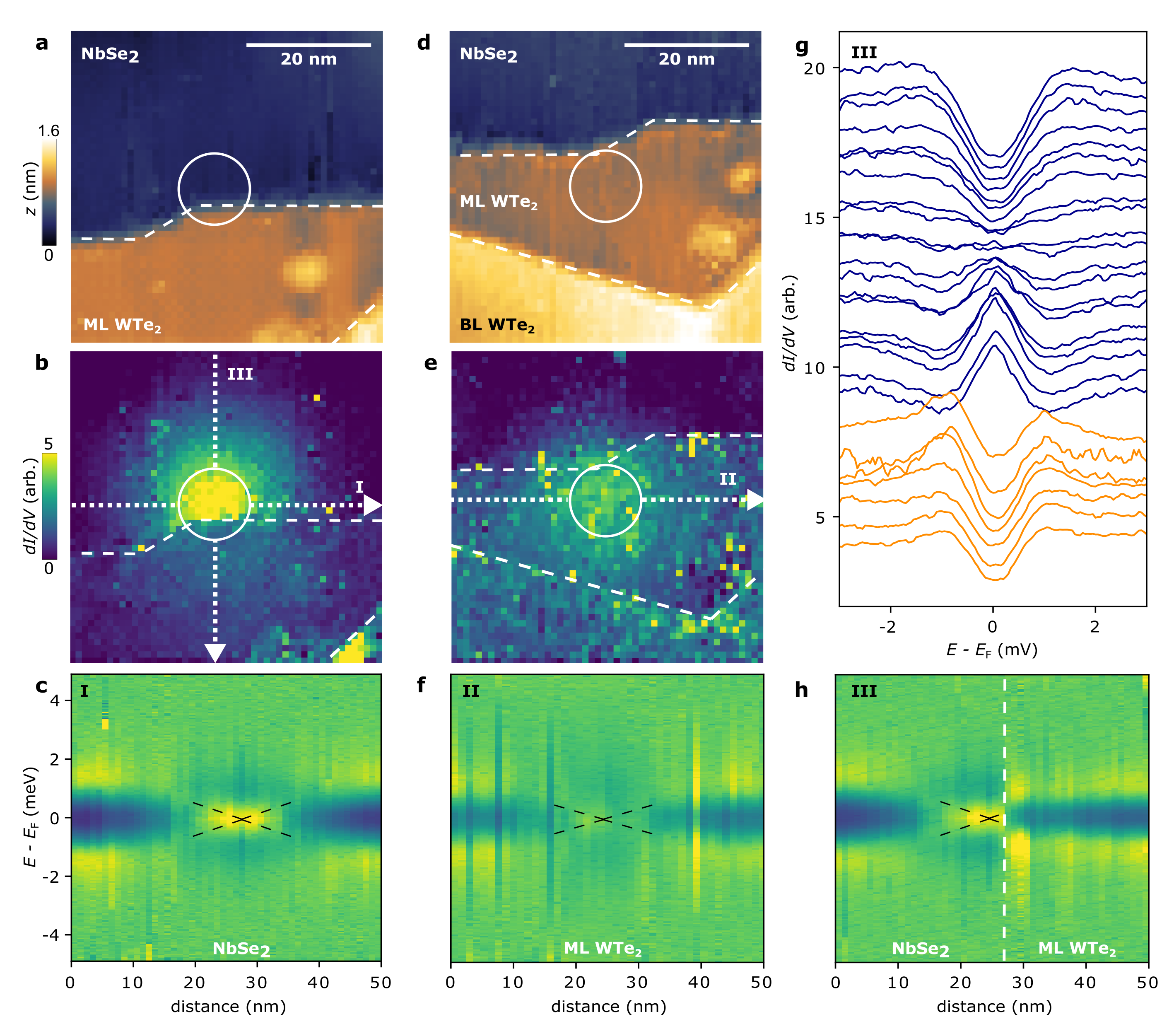}
  \caption{\label{Fig2} \textbf{Influence of an Abrikosov vortex on the induced superconductivity in monolayer \WTe.}
  \textbf{(a)} STM topography of the ML \WTe edge next to which a flux vortex, indicated by the white circle, was positioned ($B_{\rm ext}=328\,$mT). \textbf{(b)} \didv map at zero bias voltage in the same scan frame as panel a, showing the flux vortex in the \NbSe as an area of strongly increased tunnelling conductance inside the white circle. The edges of the ML \WTe region are indicated by white dashed lines in panels a and b. \textbf{(c)} \didv spectral map recorded  along arrow I parallel to the ML \WTe edge and across the flux vortex in panel b. \textbf{(d)} Topography of the ML \WTe terrace on which an Abrikosov vortex was placed at the position indicated by the white circle ($B_{\rm ext}=450\,$mT). \textbf{(e)} \didv map at zero bias voltage in the same scan frame as panel d, showing the flux vortex as an area of slightly increased tunnelling conductance inside the white circle. The edges of the ML and BL \WTe regions are indicated by white dashed lines in panels d and e. \textbf{(f)} \didv spectral map  recorded along arrow II across the vortex in panel e. \textbf{(g)} Waterfall plot of \didv spectra recorded  along arrow III across the ML \WTe step in panel b, showing an abrupt change of the spectral shape from a fully gapped spectrum on the ML \WTe (orange curves) to a peak in the \NbSe due to CdGM states (blue curves). Curves are offset by 0.5 with respect to each other. \textbf{(h)} \didv spectral map recorded along arrow III in panel b. The black dashed lines in panels c, f, and h signify the approximately linear decrease of the superconducting pairing gap towards the centre of the vortex. 
  }
\end{figure*}

\subsection{Proximity superconductivity in \WTe/\NbSe heterostructures}
Our experiments were carried out on a \WTe/\NbSe heterostructure with regions of ML  and bilayer (BL) \WTe and exposed areas of the supporting $\sim80\rm\,nm$ thick \NbSe flake (Fig.~\ref{Fig1}a, b).  It was fabricated by a modified dry-transfer flip technique \cite{lupke2020} (Methods section), which ensures that during the stacking neither the surface nor the interface come into contact with polymers or solvents. This approach is crucially important here, because scanning tunnelling spectroscopy (STS) experiments are extremely sensitive to the inevitable contamination caused by such contacts.
 
The evolution of the superconducting pairing gap $\Delta$ as a function of \WTe thickness shows a step-wise shrinking (Fig.~\ref{Fig1}c, d), which is in line with previous reports \cite{lupke2020, Weber2022}. It indicates the expected weakening of the proximity-induced superconductivity in the \WTe.  However, at the ML \WTe/\NbSe step we observe an additional constriction of the gap (Fig.~\ref{Fig1}d) on the upper terrace next to the edge of the topologically nontrivial ML \WTe ---at precisely the position of the QSH edge state, known to occur on top of the step edge \cite{Jia2017,Crommie2017,lupke2020,Lupke2022,Weber2022}. A similar but weaker constriction is also seen at the BL \WTe/ML \WTe step. In this case it occurs below the trivial BL \WTe terrace (Fig.~\ref{Fig1}d), i.e., again at the expected position of the QSH edge state \cite{Lupke2022}. In contrast, such a constriction is not observed at the BL \WTe/\NbSe step (Extended Data Fig.~\ref{EDFig1}). At both step edges of the ML \WTe region we thus observe induced pairing gaps $\Delta$ that differ from the one on the adjacent 2D terrace. This points to an intrinsically different pairing mechanism in the QSH edge state.
In the following, we will focus on the edge state at the ML \WTe/\NbSe boundary, since it is easier to access by the STM tip \cite{Lupke2022}.

\subsection{Caroli-de Gennes-Matricon states}
Next, we apply an external magnetic field perpendicular to the surface  to induce Abrikosov flux vortices in the heterostructure. By carefully tuning the magnetic field, we can position vortices in various regions of interest, with Fig.~\ref{Fig2} displaying two representative situations: a vortex located in the bare \NbSe and just intersecting the step edge (Fig.~\ref{Fig2}a-c), and a vortex located entirely in the ML \WTe (Fig.~\ref{Fig2}d-f). The properties of a vortex in the bare \NbSe are revealed by a horizontal cut along arrow I in Fig.~\ref{Fig2}b. Coming from outside the vortex, we observe the expected gradual closing of the pairing gap, with a peak at the Fermi energy emerging in the vortex centre (Fig.~\ref{Fig2}c) \cite{Hess1989}. This well-known feature derives from Caroli-de Gennes-Matricon (CdGM) states \cite{caroli1964}. They are bound quasiparticles of energy $E = \pm\mu\Delta^2/E_{\rm F}$, where $\mu=(\frac{1}{2}, \frac{3}{2}, ...)$ is the orbital angular momentum and $\Delta$ the intrinsic pairing gap in the absence of magnetic field. The bound states appear in the confinement potential for normal electrons that is created by the linearly vanishing superconducting energy gap close to the vortex core. Because in \NbSe the ratio $\Delta^2/E_{\rm F}$ is very small, we observe a single peak centered at zero bias instead of a series of states for different values of $\mu$ \cite{chen2018}. For the vortex in the proximitized ML \WTe (Fig.~\ref{Fig2}d-f), the same qualitative picture emerges as for the vortex in the \NbSe, as revealed by in Fig.~\ref{Fig2}e, f (see also Extended Data Fig. \ref{EDFig7}). Quantitative differences are readily explained by the fact that in STM we now observe the CdGM states in the ML \WTe, where the induced pairing gap $\Delta$, and therefore $\epsilon$, is smaller than in the bare \NbSe.

Coming back to the vortex intersecting the ML \WTe edge (Fig.~\ref{Fig2}a), spectra taken along arrow III in Fig.~\ref{Fig2}b elucidate the interaction of the flux vortex in the \NbSe with the proximitized \WTe and its QSH edge state. Remarkably, we observe an extremely sharp transition in which the \didv spectrum changes from a fully developed CdGM peak in \NbSe to a well-formed superconducting gap in the ML \WTe edge state (Fig.~\ref{Fig2}g, h). This occurs from one pixel to the next, corresponding to a lateral distance of less than $10\,$\AA, while on the opposite side, in the \NbSe, the transition from peak to gap is gradual as expected (Fig.~\ref{Fig2}g, h). 
The gapped spectrum in ML \WTe demonstrates the absence of CdGM states and thus also the lack of any appreciable ring current there. However, in the \NbSe the ring current must extend below the ML \WTe, because flux vortices are always fully encircled by a ring current with a total phase accumulation of $2\pi$. We can thus conclude that in this configuration the vortex does not extend into the ML \WTe, in contrast to the situation in Fig.~\ref{Fig2}d-f.

\subsection{Probing the edge-state superconductivity} 
The special experimental situation where a flux vortex is extending below the edge of the ML \WTe can be employed to probe how the two-dimensional \WTe and its one-dimensional QSH edge state respond to the laterally varying \NbSe pairing gap underneath and the laterally varying penetrating magnetic field. To evaluate the local strength of the superconductivity across the surface, we analyse the relative depth of the pairing gap, defined by 
\begin{eqnarray}\label{eq:1}
    \delta_{\mathrm{rel}}\equiv\frac{dI/dV ({\rm 3\,mV}) - dI/dV({\rm 0\,mV})}{dI/dV({\rm 3\,mV})}.
\end{eqnarray}
A value $0 < \delta_{\mathrm{rel}}< 1 $ indicates a superconducting gap (with $\delta_{\mathrm{rel}}=1$ indicating a fully developed gap), while $\delta_{\mathrm{rel}}<0$ indicates the presence of a peak at zero bias (such as stemming from CdGM states). Although $\delta_{\mathrm{rel}}>0$ quantifies the relative \textit{depth} of the pairing gap, it also correlates well with the \textit{width} of the gap, as is evident in Fig.~\ref{Fig1}c.

\begin{figure}[!htbp]
  \centering
  \includegraphics[width=63mm]{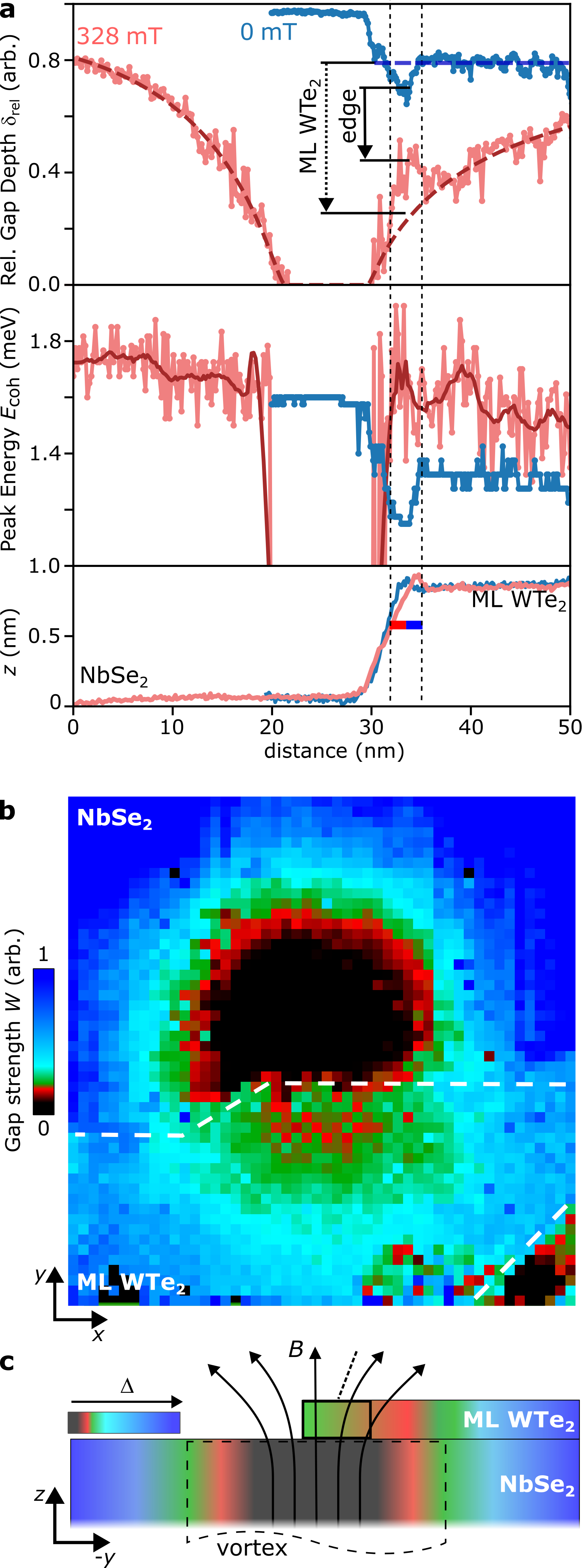}
  \caption{\label{Fig3} \textbf{Robustness of the superconducting QSH edge state.}
  \textbf{(a)} Top panel: Relative depth $\delta_{\rm rel}$ of the pairing gap, evaluated along a line across the ML \WTe/\NbSe step edge, with (red) and without (blue) vortex.
  Blue and red dashed lines describe a constant ($0\rm\,mT$) and a log-normal distribution (328\rm\,mT), respectively. Vertical arrows indicate the weakening of $\delta_{\rm rel}$ by the vortex, for the ML \WTe and QSH edge state, respectively.
  Middle panel: FD-STS of the coherence peak energy $E_{\rm coh}$ along the same line. Bottom panel: Corresponding topography cross section. Red and blue symbols and the vertical dashed lines indicate the location of the QSH edge state. 
  $\delta_{\rm rel}$ and  $E_{\rm coh}$ are calculated from the data shown in Fig. \ref{Fig1}d and Extended Data Fig.~\ref{EDFig8}.
  \textbf{(b)} Colour map of the gap strength $W$ at $V=0$, determined by FD-STS (Methods), in the same scan frame as Fig.~\ref{Fig2}a, showing a channel of stronger superconductivity along the \WTe step edge. The edges of the ML \WTe region are indicated by white dashed lines. \textbf{(c)} Schematic vertical cross section through the vortex in panel b. The vortex, indicated by the dashed black line, threads the \NbSe flake and extends under the ML \WTe and its QSH edge state, the latter outlined by a black box, but in this configuration does not thread the ML \WTe. Arrows indicate the magnetic field penetrating the flux vortex and \WTe. The colour scale represents the  pairing gap $\Delta$.
   }
\end{figure}

Plotting $\delta_{\mathrm{rel}}$ at $B_{\rm ext}=0$  along a line across the ML \WTe edge (dotted arrow in Fig.~\ref{Fig1}b) reveals a reduction of approximately 12\%\ (null hypothesis probability $p<0.005$, see Methods) in the QSH edge state compared to the 2D interior of the ML \WTe (blue curve in the top panel of Fig.~\ref{Fig3}a), in full agreement with the observations discussed in the context of Fig.~\ref{Fig1}d. Remarkably, in the presence of the vortex in Fig.~\ref{Fig2}a, b this situation is reversed. Now, the superconductivity in the QSH edge state is locally the strongest, as the red curve in the top panel of Fig.~\ref{Fig3}a demonstrates: Starting from the \NbSe, $\delta_{\mathrm{rel}}$ decreases towards the centre of the vortex, becoming negative in its core (CdGM states) before rapidly increasing back to positive values at the ML \WTe/\NbSe step edge---this is connected to the abrupt change in spectral shape as observed in Fig.~\ref{Fig2}g---and reaching a local maximum in the QSH edge state. Beyond the QSH edge state, $\delta_{\mathrm{rel}}$ decreases in the ML \WTe, before finally rising again. 
Thus, in a region that is about $3\rm\,nm$ wide, $\delta_{\mathrm{rel}}$ is \textit{locally increased} by approximately 18\%\ ($p<0.004$) over the extrapolated behavior of the ML \WTe (dashed red line in the top panel of Fig.~\ref{Fig3}a). This region coincides with the region of \textit{locally decreased} $\delta_{\mathrm{rel}}$ in the absence of a magnetic field which marks the QSH edge state.

Analysing the energy positions of the superconducting coherence peaks $E_{\rm coh}$ (Fig.~\ref{Fig3}a, middle panel) with a feature-detection scanning tunnelling spectroscopy (FD-STS) algorithm (Methods and Refs.~\cite{Sabitova2018,Martinez-castro2022}) supports the observation of a distinct behaviour of the QSH edge state: When the vortex is absent, the coherence peaks in the edge state are approximately $10\%$ closer to $E_{\rm F}$ ($p<0.003$) than in the ML \WTe (Fig.~\ref{Fig3}a, middle panel, blue curve). When the vortex is present, the situation is not as clear, as, due to the weak intensities of the coherence peaks, their exact energy is more difficult to determine. Nevertheless, we observe a trend to higher $E_{\rm coh}$, i.e. larger gap sizes, in the region of the QSH edge state (Fig.~\ref{Fig3}a, middle panel, red curve).

Taking all the evidence together, we conclude that the superconductivity in the QSH edge state at zero magnetic field is weaker than in its 2D surrounding, but at the same time more robust against the presence of the flux vortex underneath and its locally increased magnetic field. Clearly, these distinctive properties indicate an unconventional type of superconductivity, different from the one in the 2D interior of the ML \WTe. Since an enhanced robustness against magnetic fields is a hallmark of the triplet-like pairing in the spin-momentum-locked QSH edge state, our data therefore provide direct experimental proof for 1D TSC in the proximitized QSH edge state. 

\begin{figure}[!htbp]
  \centering
  \includegraphics[width=80mm]{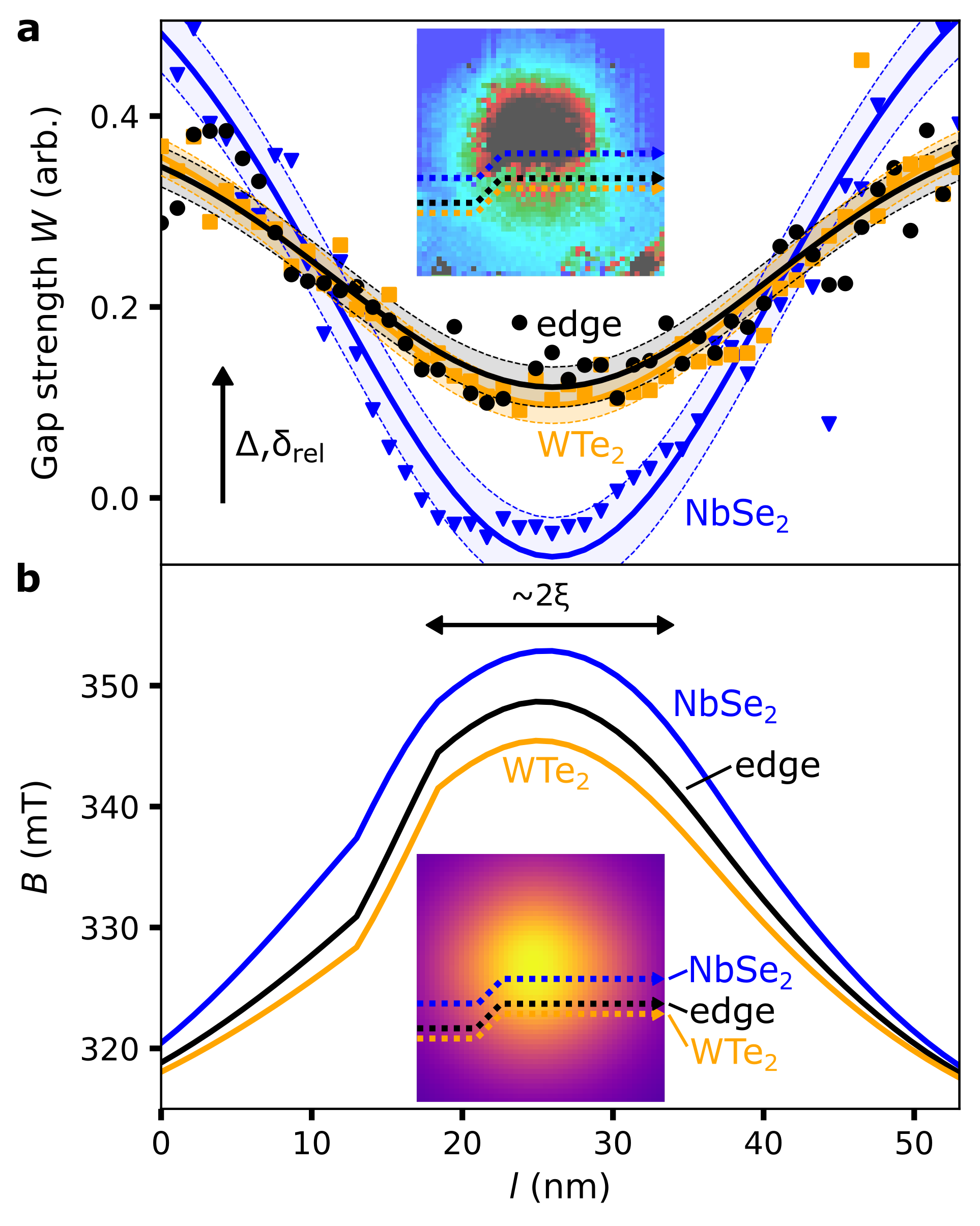}
  \caption{\label{Fig4} \textbf{Lateral proximity effect in the QSH edge state.}
  \textbf{(a)} FD-STS gap strength $W$ as function of distance $l$ along three paths parallel to the step edge, directly at the position of the edge (black circles), inside the ML \WTe (orange squares, 2\,nm from the step edge) and on the \NbSe (blue triangles, 5\,nm from the  step edge).  The solid lines are Gaussian fits to the data points, with dashed lines and shaded areas marking $1\sigma$ confidence intervals, respectively (Methods). The black arrow indicates the direction of increasing pairing gap $\Delta$ and relative gap depth $\delta_{\rm rel}$. Inset: FD-STS map from Fig. \ref{Fig3}b, with colour-coded paths. \textbf{(b)} Calculated magnetic field of the \NbSe flux vortex, plotted along the three paths defined in panel a. Note the different order of ML \WTe (orange curves) and QSH edge state (black curves) in the calculated magnetic fields and the FD-STS intensities, respectively. Inset: Map of the magnetic field with indicated paths.
   }
\end{figure}

\subsection{Lateral self-proximity effect in the 1D topological superconductor} 
Having demonstrated the side-by-side coexistence of different types of superconductivity in the heterostructure, we now analyse their lateral and vertical interactions. For this purpose, we use the FD-STS algorithm to extract the steepness of the superconducting gap $W$ (Methods), which scales with both $\delta_{\rm rel}$ and $E_{\rm coh}$, but is a more sensitive measure of the gap strength. Extracting $W$ from the \didv map in Fig. \ref{Fig2}b again reveals the sharp truncation of the vortex's signature at the step edge (Fig.~\ref{Fig3}b). However, in contrast to the \didv map, the $W$ map shows a vortex-related structure on the ML \WTe (green and red circle segments in Fig.~\ref{Fig3}b), which agrees well with the single line-cut of $\delta_{\rm rel}$ in Fig.~\ref{Fig3}a.
This structure must stem from the interaction of the ML \WTe with the vortex in the \NbSe, which extends seamlessly underneath the \WTe flake (see above).

Remarkably, within the region of the ML \WTe in Fig.~\ref{Fig3}b that is affected by the vortex, the red segment is separated from the step edge by a green channel. This channel of enhanced superconductivity close to the vortex core 
precisely tracks the QSH edge state, also following the kink in the step edge. 
It is important to note that,  since the proximity effect from \NbSe acts vertically, the gap distribution in the proximitized ML \WTe should in principle directly correlate with the pairing gap in the underlying \NbSe. Since this is not the case here, we suggest the existence of a lateral self-proximity effect.

To analyse this effect in detail, we plot the gap strength $W$ along three parallel paths, one tracing the QSH edge state, the other two offset by $5\rm\,nm$ into the \NbSe and $2\rm\,nm$ into the ML \WTe (Fig.~\ref{Fig4}a). Using the coherence length $\xi$ and penetration depth $\lambda$ of \NbSe, we calculate the magnetic field in the \NbSe vortex along these paths (Fig.~\ref{Fig4}b, Methods). While the penetrating field in the centre of the vortex reaches a maximum of $B\sim 352\rm\,mT$, the field is smaller in the QSH edge state and the ML \WTe, because they are further away from the vortex core. At the same time Fig.~\ref{Fig4}a reveals that the QSH edge state as well as the ML \WTe exhibit superconductivity that does not reflect the vanishing gap in the \NbSe below. Because the ML \WTe and its QSH edge state are not intrinsically superconducting and the vertical proximity effect essentially drops out as origin of their superconductivity, lateral self-proximity effects from outside the vortex within both must therefore be at play. 
Yet, despite the higher field in the QSH edge state we observe a significantly stronger superconductivity there ($p<0.17$) than in the ML \WTe (black vs. orange curves in Fig.~\ref{Fig4}a).
This again points to distinct superconductivities in the ML \WTe and QSH edge state, respectively.
Notably, the lateral self-proximity effect in the QSH edge state is approximately twice as strong as in the ML \WTe:
In the absence of the QSH edge state, $\delta_{\mathrm{rel}}$ at the edge of the ML \WTe would be $\sim75\%$ smaller than without flux vortex (dotted arrow in Fig.~\ref{Fig3}a). In contrast, for the edge state the reduction in the  presence of the vortex is only $\sim30\%$ (solid arrow in Fig.~\ref{Fig3}a). We explain this with the topological protection in the QSH edge state.  

For the lateral self-proximity effect to work effectively, the mean free path in the QSH edge state must be equal or larger than the radius of the Abrikosov vortex ($\sim 2\xi\approx15\,$nm). It should be noted, however, that even in the presence of the lateral self-proximity effect, the Cooper pairs in the QSH edge state are exposed to the penetrating magnetic field from the vortex underneath, which leads to a local weakening of the pairing gap in the QSH edge state as the vortex is crossed. This can clearly be observed in Fig.~\ref{Fig4}a. 

In conclusion, we demonstrated that individual Abrikosov flux vortices can be used to probe superconducting properties locally, on length scales substantially below the vortex diameter. With this approach we showed that the QSH edge state of ML \WTe is topologically superconducting, exhibiting a notable robustness against local magnetic fields and kinks in the edge, as well as a strong lateral poximity effect. We anticipate that our result marks a crucial step towards the engineering of Majorana bound states based on QSH insulators \cite{Alicea2012}.

\section{Methods}
\subsection{Sample fabrication}
\WTe and \NbSe were exfoliated onto $285\rm\,nm$ SiO$_2$/Si substrates and were assembled into heterostructures using a modified dry-transfer flip technique \cite{lupke2020}. In short, the flakes were picked up in reverse order, the heterostructure was then flipped upside down and placed onto pre-evaporated Ti/Au leads on a separate substrate, which was subsequently mounted to a standard STM sample plate. Samples were fabricated in an argon-filled glovebox and transferred to the STM chamber using an ultra-high vacuum suitcase ($p\sim10^{-10}\rm\,mbar$), i.e., without any exposure to air.

\subsection{Scanning tunnelling microscopy/spectroscopy}
Scanning tunnelling data were acquired at the Centre of Low Temperature Physics in Košice in ultra-high vacuum at a base pressure of $p<10^{-10}\rm\,mbar$ and a base temperature of $1.14\rm\,K$ using a mechanically cut Au tip. The set point parameters were $V = 5\rm\,mV$, $I = 300\rm\,pA$ for all STM/STS measurements, except those reported in Fig.~\ref{Fig1}b, which was recorded at $V = 472\rm\,mV$, $I = 20\rm\,pA$. Tunnelling spectra were acquired using standard lock-in techniques at $f=789 \rm\,Hz$ and $V_{\rm mod}=100\rm\,$\textmu V.
The \didv maps in Fig. \ref{Fig2}a, b, d, e are recorded in $50\times 50$ arrays, over areas of $50\,\mathrm{nm}\times 50\,\mathrm{nm}$. 

The effective electronic temperature of the tunnelling junction was $T_{\rm eff}=2.16\rm\,$K, as determined by fitting the superconducting gap of \NbSe. Despite recent evidence for a spatial superconducting gap anisotropy of \NbSe\cite{sanna2022},
we used a simplified isotropic two-gap model, which is sufficient for our purposes. Here the density of states is given by
\begin{align}
    N_\mathrm{2g}(E,\Delta_1,\Delta_2)=CN(E,\Delta_1)+(1-C)N(E,\Delta_2),
\end{align}
with the density of states 
\begin{align}
    N(E,\Delta)=N_0\Re\Big[\frac{E}{\sqrt{E^2-\Delta^2}}\Big]
\end{align}
from standard BCS theory, where $\Delta$ is the superconducting gap and $N_0$ the normal density of states. To extract the effective tip temperature $T_\mathrm{eff}$, the finite temperature differential conductance is derived as the convolution of the BCS density of states with the derivative of the Fermi function,
\begin{align}
    \frac{dI}{dV}(eV)\propto \int_{-\infty}^{\infty}dE\frac{e^{\frac{E-eV}{k_\mathrm{B}T_\mathrm{eff}}}}{\big[ e^{\frac{E-eV}{k_\mathrm{B}T_\mathrm{eff}}}+1 \big]^2}N_\mathrm{2g}(E,\Delta_1,\Delta_2).
\end{align}
We fitted this function to the \NbSe data by varying $\Delta_1$, $\Delta_2$, $C$ and $T_\mathrm{eff}$. 
The resulting gap values ($\Delta_1=1.36$\,meV and $\Delta_1=1.0$\,meV) are in agreement the literature \cite{Weber2022}.
We have applied the same model to extract the superconducting gaps in the \WTe spectra, where we fixed $T_\mathrm{eff}$ to the value obtained from fitting \NbSe and only varied $\Delta_1$,  $\Delta_2$ and $C$. The resulting curves and parameters are displayed in Extended Data Fig. \ref{EDFig2}a-c. 

\subsection{Feature detection-scanning tunnelling spectroscopy}
FD-STS was employed for an unbiased detection of peaks and dips in the \didv spectra \cite{Martinez-castro2022,Sabitova2018}. Specifically, we applied FD-STS to detect the dips of the superconducting gaps and the peaks of the CdGM states or the coherence peaks.
The method is visualized in Extended Data Fig.~\ref{EDFig3} and works as follows: (1) In a first step, the intrinsic noise in each of the \didv spectra was smoothed out with a Savitzky-Golay filter. To avoid filtering out meaningful signal, we used a sliding window of 350~\textmu$V$. The Savitzky–Golay filter does not only provide the smoothed $dI/dV$, but also the smoothed second derivative $d^2I/dV^2$. (2) Next, we identified peaks and dips by finding zero crossings in $d^2I/dV^2$. (3) We then assigned a weight $W$ to the feature detected in this way by determining the directed difference between each feature's two next-lying extremal values of $d^2I/dV^2$, where peaks yield $W>0$, dips $W<0$ (Extended Data Fig.~\ref{EDFig3}c). When FD-STS is applied to the superconducting gap, $W$ quantifies the steepness of the gap and is therefore referred to as gap strength $W$ in the main text. Similarly, application to the coherence peaks leads to the coherence peak strength $W$. (4) Finally, the weights $W$ of the features were plotted as a map. 
Features for which the algorithm did not return a result were set to zero. This concerns only a few data points on \NbSe, where CdGM states result in $d^2I/dV^2$ being too noisy.

We validated the FD-STS map of the superconducting gap strength $W$ in Fig.~\ref{Fig3}b by comparing it to the relative depth $\delta_{\mathrm{rel}}$ as defined in Eq.~\ref{eq:1}. Plotting the FD-STS gap strength $W$ along the same lines as $\delta_{\mathrm{rel}}$ in Fig.~\ref{Fig3}a, we find an almost identical behaviour, as shown in Extended Data Fig.~\ref{EDFig4}a. Moreover, the FD-STS map in Extended Data Fig.~\ref{EDFig4}b, which was generated from the energies of the coherence peaks, shows essentially the same behaviour as the FD-STS map of the superconducting gap strength $W$ in Fig.~\ref{Fig3}b.

The Gaussian fits to line FD-STS line profiles displayed in Fig.~\ref{Fig4}a are given by the equation
\begin{eqnarray}
    y = A\cdot\exp\left({\frac{(l-l_0)^2}{2\sigma^2}}\right)+B,
\end{eqnarray}
where $l_0$ was fixed to the centre of the vortex and $\sigma=2\xi$. Parameters $A$ and $B$ were used to fit the data points.

\subsection{Vortex calculations}
Following Ref.~\cite{tinkham}, the magnetic field around a single vortex was calculated in radial coordinates as 
\begin{align}
    B(r)={\frac {\Phi_{0}}{2\pi \lambda ^{2}}}K_{0}\left({\frac {r}{\lambda }}\right),
    \label{eq:B_radial}
\end{align}
where $K_0$ is the zeroth-order modified Bessel function of the second kind, and $\Phi_{0}$ is the flux quantum. Inside the vortex ($r\lesssim\xi$), the field was approximated by
\begin{align}
    B(0) = {\frac {\Phi _{0}}{2\pi \lambda ^{2}}}\ln(\frac{\lambda}{\xi}).
        \label{eq:B_core}
\end{align}
We used $\lambda=69\rm\,nm$ as the penetration depth and $\xi=7.7\rm\,nm$ as the coherence length of \NbSe \cite{de1973anisotropy}. To get a continuous distribution of the $B$ field, we interpolated between the two solutions (Eqs.~\ref{eq:B_radial} and \ref{eq:B_core}) using a two-dimensional spline with a smoothing factor, which approximately corresponds to a gliding average with window size $\sim\xi$. To simulate our experiments, we constructed a discrete triangular vortex lattice with a density corresponding to the flux density of the externally applied magnetic field ($a=1.075\sqrt{\Phi_0/B}\approx85\rm\,nm$, which corresponds to an intermediate flux density \cite{tinkham}). The fields penetrating the individual vortices are then added up and evaluated around a single vortex in the centre of the lattice (inset in Fig.~\ref{Fig4}b).

\subsection{Statistical analysis}
To determine the statistical significance of the data measured at the edge of the ML \WTe (data in Fig.~\ref{Fig3}a) we performed a Student's $t$-test against the null hypothesis \cite{Student1908}. For this, we determined the standard deviation of the data without the edge signal from the regression curve assuming a normal (Gaussian) distribution. To avoid underestimating the probability $p$ that the observed edge signal is due to random noise, we calculated $p$ using the cumulative Student's $t$ distribution function, which not only takes into account the signal's distance from the regression curve in units of $\sigma$, but also the relatively small sample size of only $n=10$ to $15$ data points, for which the Gauss error function would underestimate $p$. Using the cumulative Student's $t$ distribution, we find small $p$ values which confirm our claims.

\section{Acknowledgements}
The authors acknowledge helpful discussions with Samir Lounis and thank Francois C. Bocquet for technical support.  Furthermore, we are grateful to the Helmholtz Nano Facility for its support regarding sample fabrication. The authors acknowledge funding from the European Union’s Horizon 2020 Research and Innovation Programme under Grant Agreement no 824109 (European Microkelvin Platform). J.M.C., T.W., K.J., M.T. and F.L. acknowledge funding by the Deutsche Forschungsgemeinschaft (DFG, German Research Foundation) within the Priority Programme SPP 2244 (project nos. 443416235 and 422707584). J.M.C., Z.L., F.S.T. and F.L. acknowledge funding from  the Bavarian Ministry of Economic Affairs, Regional Development and Energy within Bavaria’s High-Tech Agenda Project ''Bausteine f\"ur das Quantencomputing auf Basis topologischer Materialien mit experimentellen und theoretischen Ans\"atzen''.  J.M.C. acknowledges funding from the Alexander von Humboldt Foundation. T.S., P.S. and O.O. acknowledge the support of APVV-20-0425, VEGA 2/0058/20, Slovak Academy of Sciences project IMPULZ IM-2021-42, COST action CA21144 (SUPERQUMAP) and EU ERDF (European regional development fund) Grant No. VA SR ITMS2014+ 313011W856. M.T. acknowledges support from the Heisenberg Program (Grant No. TE 833/2-1) of the German Research Foundation. F.L. acknowledges financial support by Germany’s Excellence Strategy - Cluster of Excellence Matter and Light for Quantum Computing (ML4Q) through an Independence Grant. J.Q.Y. was supported by the US Department of Energy, Office of Science, Basic Energy Sciences, Materials Sciences and Engineering Division.

\vspace{3mm}

\paragraph{Author contributions.}
J.M.C., F.S.T., M.T. and F.L. conceived the research.
J.M.C., T.W. and F.L. designed the experiments.
J.Y. grew WTe$_2$ crystals.
J.M.C., T.W., K.J. and Z.L. fabricated the samples.
T.S., O.O. and P.S. set up and provided the STM.
J.M.C. and T.W. acquired the data.
J.M.C., T.W., F.S.T., M.T. and F.L. analyzed the data. 
J.M.C., T.W., P.S.,  M.T., F.S.T. and F.L. developed the physical interpretation. 
J.M.C., T.W., F.S.T and F.L. wrote the paper.
All authors commented on the manuscript.
J.M.C., F.S.T., MT. and F.L. supervised the research.

\vspace{3mm}

\paragraph{Competing financial interests.}
The authors declare no competing financial interests.

\bibliography{Refs.bib}

\setcounter{figure}{0}
\renewcommand{\figurename}{Extended Data Fig.}

\newpage
\begin{figure*}[!htbp]
  \centering
  \includegraphics[width=.66\textwidth]{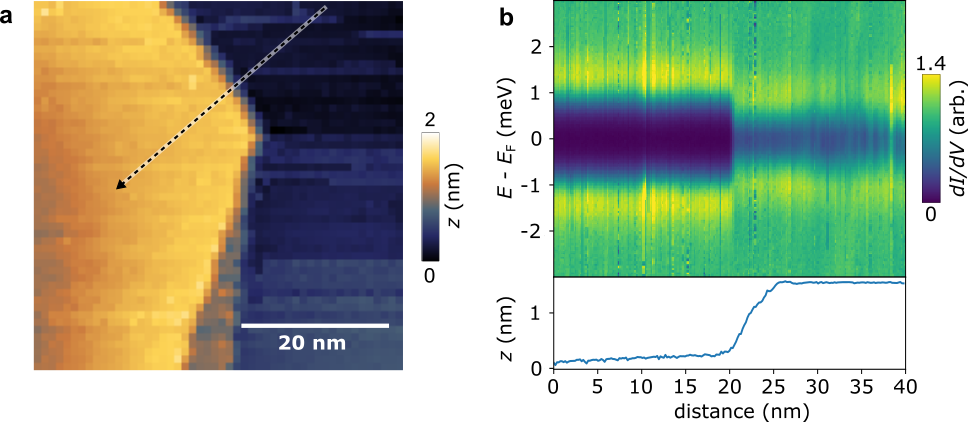}
  \caption{\label{EDFig1} \textbf{BL \WTe/ \NbSe step edge.}
  \textbf{(a)} STM topography. \textbf{(b)} \didv spectral map (top) recorded along the dashed arrow in panel a  and corresponding topography cross section (bottom).
  }
\end{figure*}

\newpage 
\begin{figure*}[!htbp]
  \centering
  \includegraphics[width=0.35\textwidth]{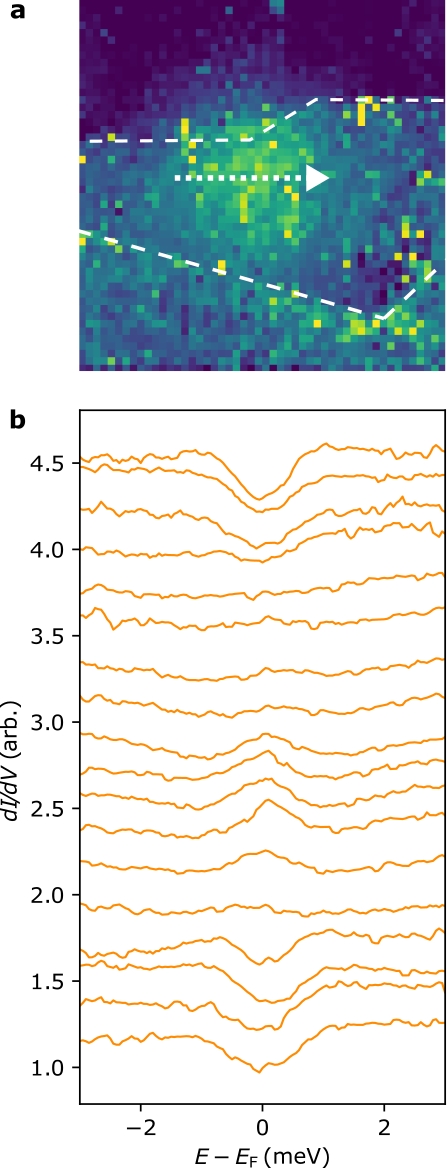}
  \caption{\label{EDFig7} \textbf{Induced Abrikosov vortex in the interior of the ML \WTe.} 
   \textbf{(a)} \didv map at zero bias voltage (same as Fig.~\ref{Fig2}e), showing the vortex on the ML \WTe. The white arrow accross the vortex indicates the line along which the spectra in panel b where extracted.
   \textbf{(b)} Waterfall plot of \didv spectra recorded along the arrow in panel a, showing the presence of CdGM states. The vertical spacing between the curves is 0.2 (arb.). 
  }
\end{figure*}

\begin{figure*}[!htbp]
  \centering
  \includegraphics[width=0.5\textwidth]{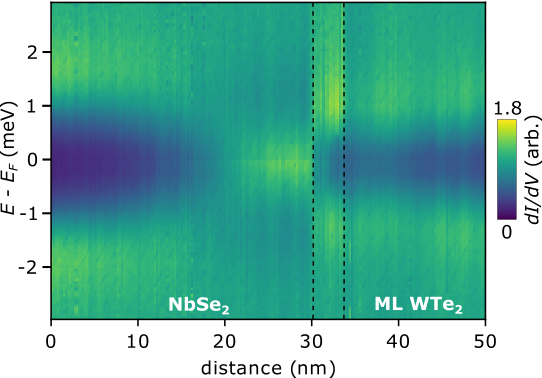}
  \caption{\label{EDFig8} \textbf{Spectral map of ML \WTe / \NbSe with Abrikosov vortex.} High-resolution spectral map taken along the dashed arrow III in Fig. \ref{Fig2}b, i.e., in the presence of the Abrikosov vortex. The edge state region is marked by dashed black lines (same as in Fig. \ref{Fig1}d). 
  }
\end{figure*}

\begin{figure*}[!htbp]
  \centering
  \includegraphics[width=1\textwidth]{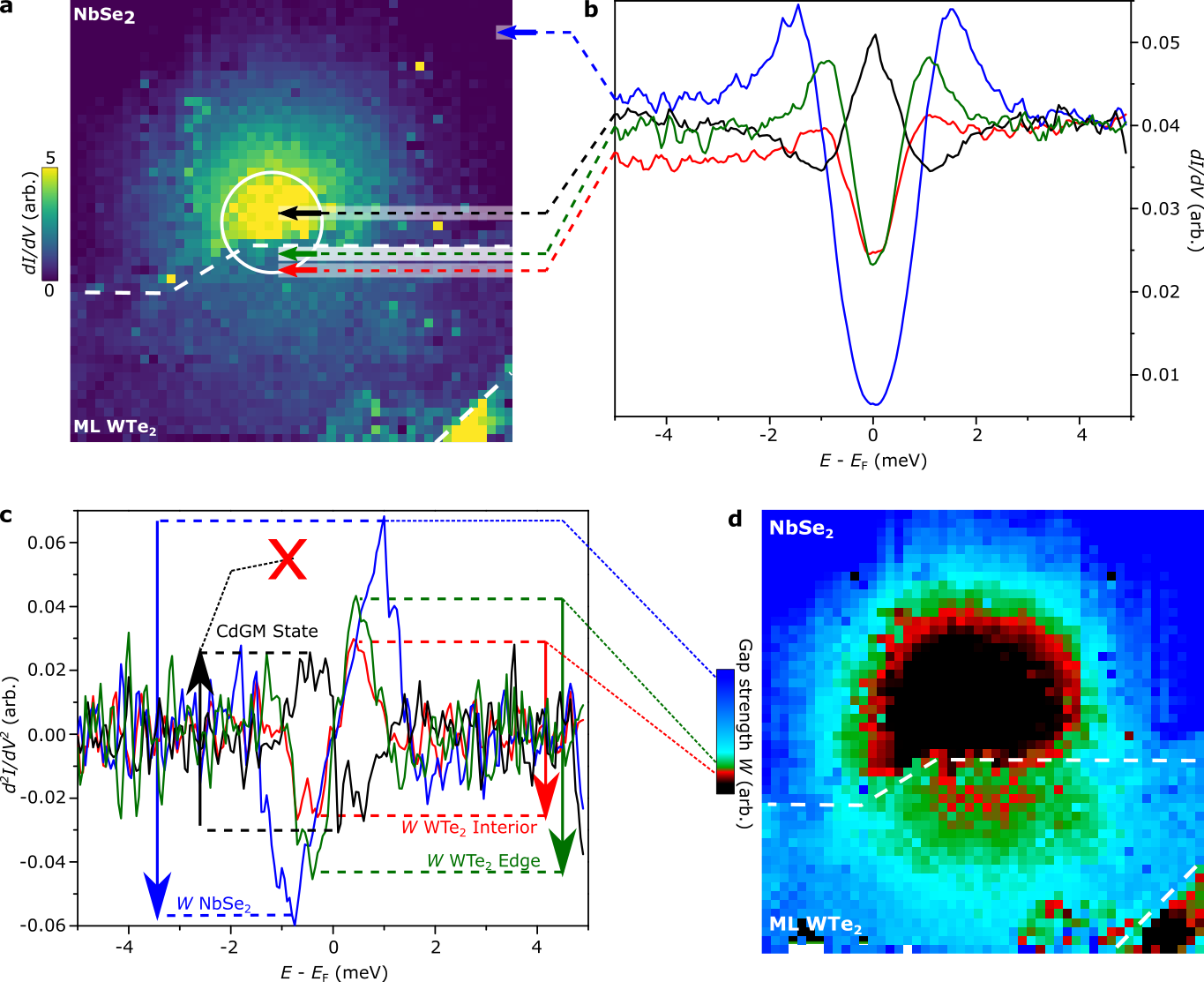}
  \caption{\label{EDFig3} \textbf{
 Illustration of the FD-STS algorithm, applied to the case of the superconducting gap strength.} 
  \textbf{(a)} \didv map at zero bias voltage (same as Fig.~\ref{Fig2}b). (b) \didv spectra taken at the positions indicated in panel a by black, blue, green and red arrows. \textbf{(c)} $d^2I/dV^2$, i.e., derivative of the \didv spectra in panel b. The weights $W$ assigned by the FD-STS algorithm are indicated by vertical arrows. Peaks in panel b yield $W>0$, dips $W<0$, corresponding to the arrow orientation. \textbf{(d)} Colour map of the extracted weights $W$, reproducing Fig.~\ref{Fig3}b.  
  }
\end{figure*}

\begin{figure*}[!htbp]
  \centering
  \includegraphics[width=1\textwidth]{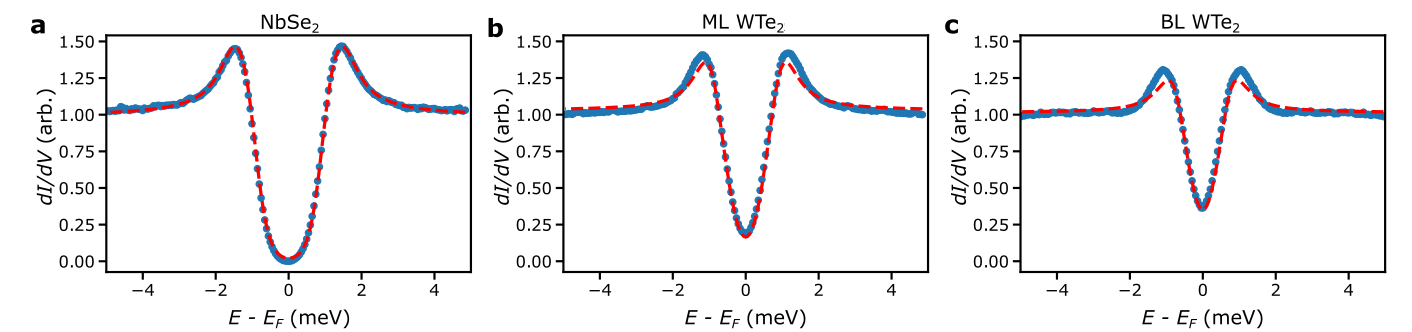}
  \caption{\label{EDFig2} \textbf{
  BCS fits to the \NbSe, ML and BL \WTe spectra in Fig.~\ref{Fig1}c.} 
  \textbf{(a)} The \NbSe fit gives an effective temperature of $T_\mathrm{eff}=(2.16\pm0.03)$\,K,  gap parameters  $(\Delta_1=1.36\pm0.02)$\,meV, $(\Delta_2=1.00\pm0.01)$\,meV and $C=0.3\pm0.03$ (for details see Methods section).
  \textbf{(b, c)} The fitted \WTe gaps (at fixed $T_\mathrm{eff}=2.16$\,K) are $\Delta_1=(0.84\pm0.02)$\,meV, $\Delta_2=(0.45\pm0.04)$\,meV, $C=0.67\pm0.06$ (monolayer) and $\Delta_1=(0.66\pm0.02)$\,meV, $\Delta_2=(0.25\pm0.06)$\,meV, $C=0.64\pm0.07$ (bilayer), respectively.   
  }
\end{figure*}
\begin{figure*}[!htbp]
  \centering
  \includegraphics[width=0.35\textwidth]{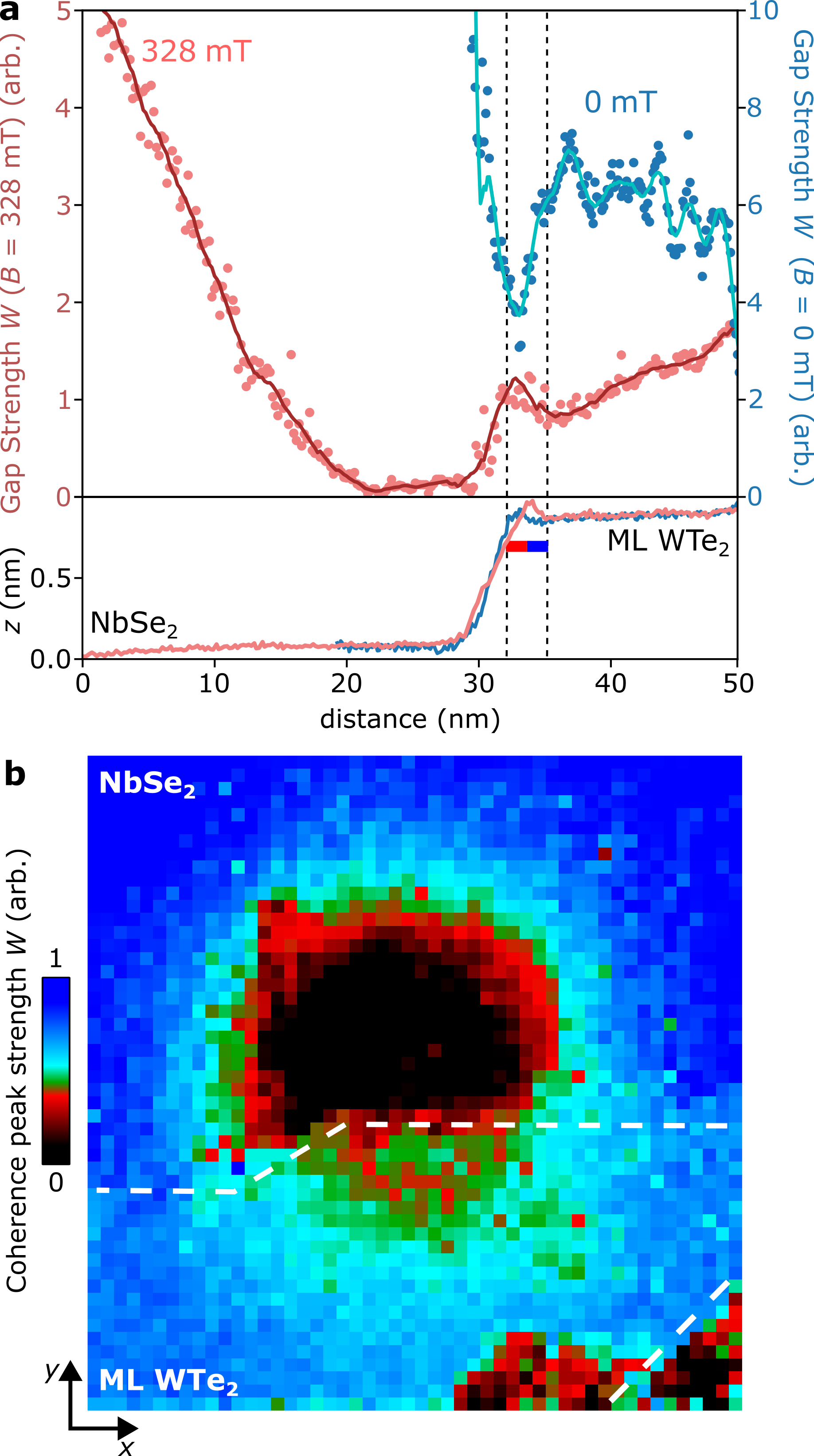}
  \caption{\label{EDFig4} \textbf{Additional data plots showing the robustness of the superconducting QSH edge state.}
  \textbf{(a)} FD-STS of the superconducting gap strength $W$, evaluated along a line across the ML  \WTe/\NbSe step edge, with (red) and without (blue) vortex. Bottom panel: Corresponding topography cross section. Red and blue symbols and the vertical dashed lines indicate the location of the QSH edge state.   
  \textbf{(b)} Colour map of the FD-STS coherence peak strength $W$, in the same scan frame as Fig.~\ref{Fig2}a. The edges of the ML \WTe region are indicated by white dashed lines.
  Both panels, a and b, essentially reproduce the features from Fig.~\ref{Fig3}, despite analysing different quantities, i.e., the gap strength and coherence peak strength, respectively.
}  
\end{figure*}

\end{document}